\documentclass[num-refs]{article}
\usepackage[margin=1in]{geometry}
\usepackage{siunitx}
\usepackage{algpseudocode} 
\usepackage{xspace}
\usepackage{csvsimple}
\usepackage{datatool}
\usepackage{booktabs}
\usepackage{xcolor}
\usepackage{amsmath}
\usepackage{graphicx}
\usepackage{multirow}
\usepackage{array}
\usepackage{algorithm}
\usepackage{caption}
\usepackage[justification=centering]{caption}
\usepackage{subcaption}
\usepackage{listings}
\usepackage{verbatim}
\usepackage{adjustbox}
\usepackage{makecell}
\usepackage{titlesec}
\usepackage{anyfontsize}
\usepackage[flushleft]{threeparttable}
\usepackage{xspace}
\usepackage{hyperref}

\newcommand{\reprozip}[0]{ReproZip\xspace}
\newcommand{\rom}[1]{\lowercase\expandafter{\romannumeral #1\relax}}

\definecolor{ao(english)}{rgb}{0.0, 0.5, 0.0}
\newcommand{\revision}[1]{#1\xspace} 

\title{An Analysis of Security Vulnerabilities in Container Images for Scientific Data Analysis}
  
\begin{document}

\author{Bhupinder Kaur, Mathieu Dugr\'e, Aiman Hanna, Tristan Glatard \\Department of Computer Science and Software Engineering\\ Concordia University\\ Montreal, Canada}


\maketitle

\begin{abstract}
Software containers greatly facilitate the deployment and reproducibility
of scientific data analyses \revision{in various platforms.} However,
container images often contain outdated or unnecessary software packages,
which increases the number of security vulnerabilities in the images,
widens the attack surface \revision{in the container host, and creates
substantial security risks for computing infrastructures at large}. This paper
presents a vulnerability analysis of container images for scientific data
analysis. We compare results obtained with four vulnerability scanners,
focusing on the use case of neuroscience data analysis, and quantifying the
effect of image update and minification on the number of vulnerabilities.
We find that container images used for neuroscience data analysis contain
hundreds of vulnerabilities, that software updates remove about two thirds
of these vulnerabilities, and that removing unused packages is also
effective. We conclude with recommendations on how to build container
images with a reduced amount of vulnerabilities.

\end{abstract}


\section{Introduction}

Software containers have emerged has an efficient solution to deploy
scientific data analyses \revision{in various platforms}, due
to their portability, ease of use, and limited overhead. 
Taking advantage of core Linux kernel features such as namespaces, control
groups\revision{,} and chroot, containers isolate processes from the host computer and \revision{often can}
control the memory, CPU, network\revision{,} and file-system resources assigned to
them. However, containers still share the kernel, mounted file systems and
some devices with the host, which raises security
concerns~\cite{martin2018docker, sultan2019container, combe2016docker} and
opens the door to privilege escalation, denial of service, information leak\revision{,}
and other types of attacks~\cite{gantikow2016providing}. 

Container images typically include full operating system (OS) distributions
in addition to data analysis software and their dependencies. They are
rarely updated due to concerns that \revision{updated} software may \revision{be
incompatible or may} interfere with the
results \revision{via numerical perturbations propagating in the
analyses in unknown ways}~\cite{gronenschild2012effects, glatard2015reproducibility}.
Images also typically include more dependencies than required, to make them
easier to reuse between experiments. As a result, over 30\% of official
images in DockerHub have been shown to contain high-priority security
vulnerabilities~\cite{gummaraju2015over}, images on average contain over
180 vulnerabilities~\cite{Shu2017}, and vulnerabilities are often caused by
outdated packages~\cite{zerouali2019relation}.

\revision{Scientific data analyses typically involve a range of computational 
infrastructures, including personal workstations, lab servers, HPC clusters, 
and cloud computing platforms. It is common for researchers to have access
 to multiple systems through a combination of credentials and to migrate
  analyses depending on their evolution. As a result, an attacker gaining 
  access to one of these systems, possibly through a vulnerable container, 
  might be able to compromise an extensive infrastructure and use it for malicious purposes.}

In this study, we focus on the vulnerabilities present in container images
\revision{used in} scientific data analysis, in particular in the
neuroimaging domain. We address the following questions:

\textit{What is the current amount of vulnerabilities in
container images \revision{used in scientific analyses}?} Vulnerabilities are possible
attack vectors that can seriously compromise the security of \revision{computing systems}
and the integrity of user data. We report vulnerability scans produced
by four popular image scanning tools: Anchore, Vuls, Clair, and Singularity tools.

\textit{Can the amount of vulnerabilities be reduced by updating the images?}  
\revision{To avoid breaking software dependencies and introducing numerical
 perturbations, container images used in scientific analyses often include
 outdated software}. We report
on the effect of software updates on the amount of
vulnerabilities found in images.

\textit{Can the amount of vulnerabilities be reduced by minifying images?} 
Container images often include more software packages than necessary for 
a typical analysis. We report on the impact of unused software packages on
the presence of vulnerabilities.

The remainder of this paper 
describes the container images and scanners used in our experiment, and our methodology for updating and minifying images.
Results present the vulnerabilities detected in container
images, quantify the effectiveness of updating and minifying images, and
explain the differences observed between scanners. In conclusion, we
provide a set of image creation guidelines for a more secure deployment of
containers \revision{in scientific analyses}.

\section{Materials and Tools}

We used container
images from two popular application frameworks, as well as
four of the major image scanners.

\subsection{Container Images}

We scanned all container images available at the time of this study on two containerization frameworks
used in neuroscience: BIDS
apps~\cite{gorgolewski2017bids} (26 images) and Boutiques~\cite{glatard2018boutiques} (18 images),
totalling
44 container images. At the time of the study, BIDS apps had 27 images,
out of which one wasn't available on DockerHub. Boutiques had 49 images,
however, only 23 unique images were listed, out of which 3 couldn't be retrieved and 2
were already included in BIDS apps. All the final 26 images
from BIDS apps were Docker images, whereas the 18 Boutiques images contained 12 Docker images
and 6 Singularity images.

\subsection{Image Scanners}

We compared the results obtained with four container image scanners: Anchore, Vuls, and
Clair to scan Docker images, and Singularity Container Tools
(Stools) to scan Singularity images. 

\href{https://github.com/anchore/anchore-engine}{Anchore} is an end-to-end, open-source container security platform. It
analyzes container images and lists vulnerable OS
packages, non-OS packages (Python, Java, Gem, and npm), and files.
In our experiments, we used Anchore Engine version 0.5.0 through Docker image \texttt{anchore/anchore-engine:v0.5.0}, and
Anchore vulnerability database version 0.0.11.

\begin{table*}
  \centering
	\footnotesize
\begin{tabular}{|c|c|c|c|}
 \hline
\textbf{OS} &	\textbf{Anchore} &	\textbf{Vuls} &	\textbf{Clair} \\
\hline
	\textbf{Alpine} & Alpine-SecDB &	Alpine-SecDB &	Alpine-SecDB \\
\hline
	\textbf{CentOS} & Red Hat OVAL Database & Red Hat OVAL Database and Red Hat Security Advisories & Red Hat Security Data \\
\hline
	\textbf{Debian} & Debian Security Bug Tracker &	Debian OVAL Database and Debian Security Bug Tracker & Debian Security Bug Tracker \\
\hline
	\textbf{Ubuntu} & Ubuntu CVE Tracker &	Ubuntu OVAL Database &	Ubuntu CVE Tracker \\
 \hline
\end{tabular}
\caption{Vulnerability databases used by scanners for different OS distributions. All scanners also refer to 
the National Vulnerability Database (NVD) for vulnerability metadata.}
\label{table:databases}
\end{table*}

\href{https://github.com/future-architect/vuls}{Vuls} is an open-source vulnerability scanner for Linux and FreeBSD. It
offers both static and dynamic scanning, and both local and remote
scanning. In our experiments, we used Vuls 0.9.0, executed through Docker image
\texttt{vuls/vuls:0.9.0} in remote dynamic mode.

\href{https://github.com/quay/clair}{Clair} is an open-source and
extensible vulnerability scanner for Docker and \revision{OCI (Open Container Initiative)} container images,
developed by CoreOS (now Container Linux), a Linux distribution to deploy
container clusters.
 We used Clair through
\href{https://github.com/arminc/clair-scanner}{Clair-scanner}, a tool to
facilitate the testing of container images against a local Clair server. 
We used Clair version 2.0.6, executed through
Docker image \texttt{arminc/clair-local-scan:v2.0.6}. For the vulnerability
database, we used Docker image \texttt{arminc/clair-db:latest}, last
updated on 2019-09-18.

\href{https://github.com/singularityhub/stools}{Singularity Tools} (Stools)
are an extension of Clair for Singularity images. Stools
exports Singularity images to \texttt{tar.gz} format, acting as a single layer Docker image
to circumvent the Docker-specific requirements in the Clair API.
In our experiments, we used Singularity Tools version 3.2.1 through Docker
image
\texttt{vanessa/stools-clair:v3.2.1}.
Since Stools uses Clair internally for scanning, the vulnerability databases used
by Stools are the same as mentioned for Clair.
To scan Singularity images, we followed the steps mentioned in the
\href{https://github.com/singularityhub/stools}{Stools documentation}.

\subsection{Vulnerability Databases}

Scanners refer to two types of
vulnerability databases (Table~\ref{table:databases}). The first one is the Open Vulnerability and
Assessment Language (OVAL) database, an international open standard that
supports various OS distributions including Ubuntu, Debian and CentOS but
not Alpine. The second one are vulnerability databases from specific OS
distributions, such as Alpine-SecDB, Debian Security Bug Tracker, Ubuntu
CVE Tracker, or Red Hat Security Data. In these databases, OS distributions often assign a
status to each vulnerability, to keep track of required and available
security fixes in different versions of the distribution. Vuls uses OVAL
databases for all distributions except Alpine. On the contrary, Clair exclusively refers to
distribution-specific databases. Anchore uses OVAL only for CentOS, as distribution-specific databases
are assumed to be more complete.
It is also worth noting that there is no vulnerability data
for Ubuntu 17.04 and 17.10 distributions in the OVAL database, since these
distributions have reached end of life, meaning that images with these
distributions cannot be scanned with Vuls.

For CentOS images, Anchore and Clair \revision{report} scanning results using Red Hat
Security Advisory (RHSA) identifiers, whereas Vuls uses the Common
Vulnerabilities and Exposures (CVE) identifiers used in OVAL. We mapped
RHSA identifiers to corresponding CVE identifiers, to allow for a
comparison between scanners.

Different vulnerabilities may be reported by scanners if scanning
experiments take place on different dates. To avoid such discrepancies, we
froze the vulnerability databases used by these scanners as of 2019-09-25.

\subsection{Image Update}

A first approach to reduce the number of vulnerabilities in container
images is to update their packages to the latest version available in the OS
distribution. To study the effect of such updates, we \revision{created} a script
(available
\href{https://github.com/big-data-lab-team/container-vulnerabilities-paper/blob/master/Scripts/update}{here})
to identify the package manager in the image, and invoke it to update all
OS packages. We updated images on 2019-11-05.

\subsection{Image Minification}

A second approach to reduce the number of vulnerabilities in the images is
to remove unnecessary packages, an operation potentially specific to each
analysis. \revision{Although vulnerabilities in unused packages could not
be directly exploited when running the analysis provided by the container
image, they still increase the risk of escalation attacks and should
therefore be avoided.}

We used the open-source \reprozip tool~\cite{rampin2016reprozip}
to capture the list of packages used by an analysis. \reprozip first
captures the list of files involved in the analysis, through system call
interception, then retrieves the list of associated software packages, by
querying the package manager. We extend this list with a passlist of
packages required for the system to function, such as \texttt{coreutils}
and \texttt{bash}, and with all the dependencies of the required packages,
retrieved using
\href{http://manpages.ubuntu.com/manpages/xenial/man1/debtree.1.html}{Debtree}.
\href{https://linux.die.net/man/1/repoquery}{Repoquery} could be used in
RPM-based distributions instead. Our minification script, available
\href{https://github.com/big-data-lab-team/container-vulnerabilities-paper/tree/master/Scripts/minification}{here},
installs \reprozip in the image to minify, runs an analysis to collect a
\reprozip trace, and finally deletes all unnecessary packages. We had used
the \href{https://github.com/ReproNim/neurodocker}{Neurodocker} tool
initially, but it did not affect the detected vulnerabilities
as it was removing unused files without using the package manager.

Image minification is a tedious operation\revision{, as it requires (1) creating
relevant analysis examples and (2) running these examples in the container
image}, to identify the packages required by the application. The
resulting \revision{minified} container image is only valid for the examples selected
in the minification process, as other executions might require a
different set of packages. 

\revision{While the Boutiques specification allows developers to 
specify analysis tests, we found that this feature was not consistently used in 
the studied container images. Therefore, we relied on analysis examples found in the 
documentation of the applications.   
Using this approach, we minified five Debian- or Ubuntu-based BIDS app images.}

\section{Results}

Figure~\ref{fig:vulnerabilities} presents our results. All the collected
data are available in our GitHub repository at
\url{https://github.com/big-data-lab-team/container-vulnerabilities-paper}
with a Jupyter notebook to regenerate the figures. 

\subsection{Detected Vulnerabilities}

\begin{figure*}
\includegraphics[width=\textwidth]{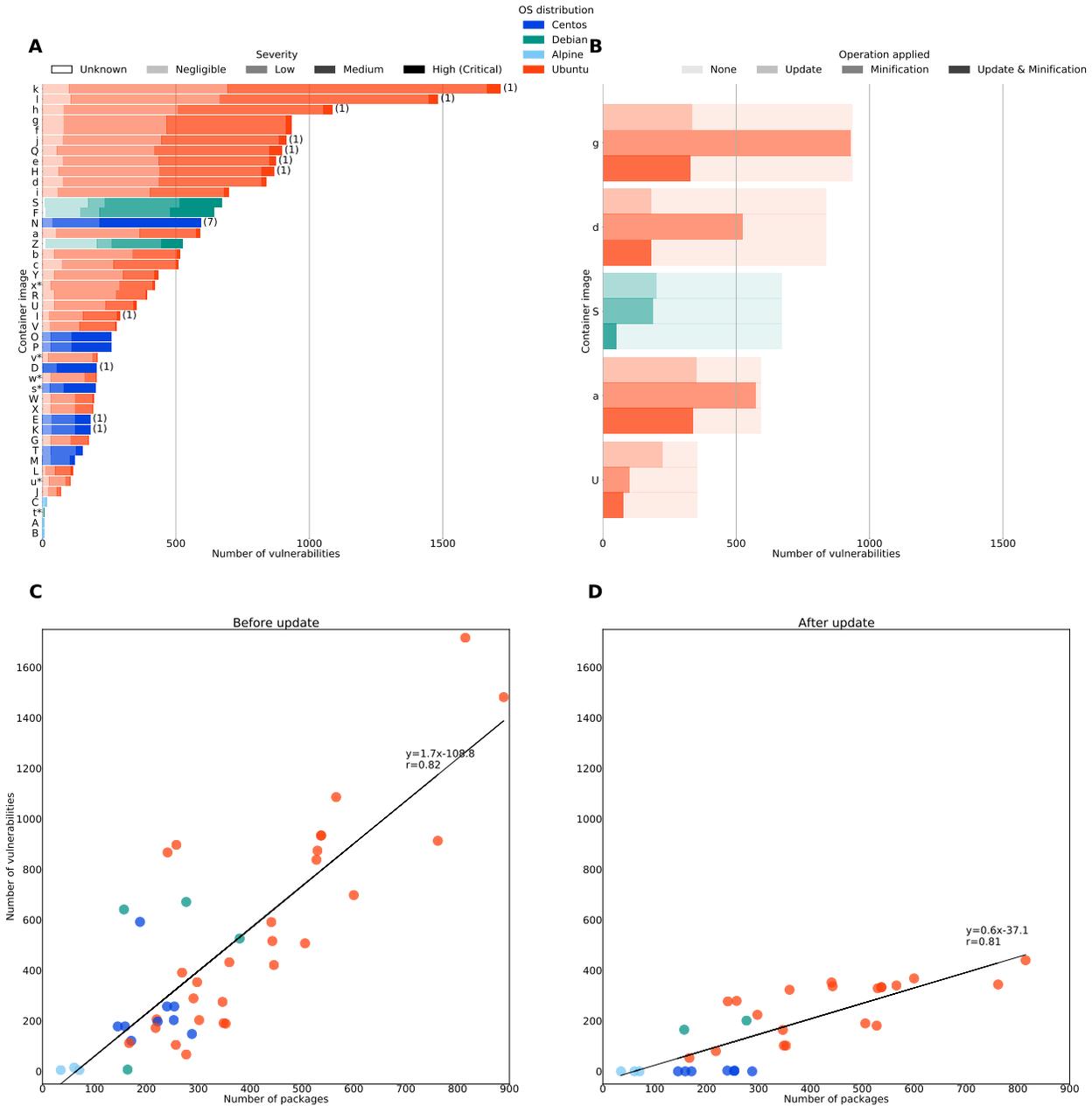}
\caption{\label{fig:vulnerabilities} Number of vulnerabilities detected by
Anchore and Stools in container images. \textbf{(A)} Number of vulnerabilities by
container image and severity, showing hundreds of detected vulnerabilities
per image. Images s*,t*,u*,v*,w* and x* are Singularity images scanned by Stools and
others are Docker images scanned using Anchore. \revision{High and critical vulnerabilities are 
represented with the same solid color and critical vulnerabilities are reported between brackets.} \textbf{(B)} Effect of image minification and
package update on 5 container images, showing that both techniques are
complementary \textbf{(C)} Number of vulnerabilities by number of
packages, showing a strong linear relationship. \textbf{(D)} Number of
vulnerabilities by number of packages \emph{after package update}, showing that software updates 
importantly reduce the number of detected vulnerabilities.}
\end{figure*}

An important amount of vulnerabilities were found in the tested container
 images (Fig~\ref{fig:vulnerabilities}-\textbf{A}), with an average of 460 vulnerabilities
  per image and a median of 321. Moreover, a significant fraction of detected vulnerabilities are
 of high severity
 (\href{https://www.first.org/cvss/specification-document}{CVSS} score
 >=7.0) and a few of them are of critical severity (CVSS >= 9.0). Remote
 attackers could possibly exploit these vulnerabilities to execute
 arbitrary code in the container \revision{or to store arbitrary files in the system}. 
 \revision{For instance, vulnerability CVE-2019-3462 could lead to remote
 code execution through human-in-the-middle attacks and vulnerability CVE-2018-1000802
 could lead to arbitrary file injections in the filesystem through unfiltered Python user input.}
 \revision{Among other consequences, remote code execution and injection of
 arbitrary files could be leveraged to steal user credentials, to use CPU
 cycles or storage space for illegitimate purposes, or to attempt
 denial-of-service attacks against the system.}
 
 Images based on the Alpine distribution 
 had the lowest numbers of vulnerabilities, but no significant difference
 in the numbers of vulnerabilities detected in 
 Ubuntu, Debian or CentOS distributions was observed.

Unsurprisingly, a strong linear relationship is found between the number of
detected vulnerabilities and the number of packages present in the
image (Fig~\ref{fig:vulnerabilities}-\textbf{C}, r=0.82,
$p\textless10^{-11}$). On average, 1.7 vulnerabilities are introduced for
each new package installation. This observation motivates a systematic
review of software dependencies by application developers, to avoid
unnecessary packages in container images. Compared to Ubuntu and Debian distributions,
CentOS images seem to have a lower number of vulnerabilities by package on
average, although data is too scarce to conclude.

\subsection{Effect of image update}

Updating container images reduces the number of vulnerabilities by package
by a factor of 3 on average, resulting in only 0.6 extra vulnerabilities by
package (Fig~\ref{fig:vulnerabilities}-\textbf{D}, r=0.81,
$p\textless10^{-7}$). Twelve container images are missing on this figure.
Six of them are Singularity images that we did not update and six of them
\revision{could not be updated by our script due to the following issues.
One image was built from base Docker image CentOS 7.1.1503 that includes
package \texttt{fakesystemd} conflicting with several other distribution
packages: updating would require either updating the base image or swapping
package \texttt{fakesystemd} for \texttt{systemd}. Three images were build
from Ubuntu 17.04 or Debian 8 that reached end of life: updading would
require changing the source list to make it point to old releases. Two
images could not be updated because some files previously installed through
the package manager were removed by other means, leading to failure of the
update process.}
\revision{Software updates did not break the tested
analyses in the remaining images. We did not investigate the potential consequences of image updates
on numerical stability as it would require a full separate study. Updating}
packages appears to be an efficient way to avoid vulnerabilities. It
is not an ultimate solution though, as a substantial number of
vulnerabilities remain.

\subsection{Effect of minification}

Using the ReproZip-based approach described previously, we minified 5
different images covering the spectrum of detected vulnerabilities
(Fig~\ref{fig:vulnerabilities}-\textbf{B}). We find that minification reduces the
number of vulnerabilities, albeit less systematically than package update.
For some container images, such as image \textbf{S}, minification removes more
than 70\% of the detected vulnerabilities. For other images, such as
image \textbf{g}, it only reduces the number of vulnerabilities by less than 1\%.
The effect of minification stems from the number of packages
that can be removed, which varies greatly across images. For
instance, images \textbf{g} and \textbf{a} have a large number of packages,
but the last majority of them is required by the analysis, which makes
minification less useful. In other cases, a limited number of unnecessary packages contain 
a significant number of vulnerabilities, which makes minification very impactful. 
This was the case in images \textbf{d}, \textbf{S} and \textbf{U}, where removing compilers
and kernel headers reduced the number of vulnerabilities by an important fraction. 
\revision{Minification did not create any errors in these images as all the
dependencies were taken care of properly. Common packages removed by the
minification process were compilers, unused Python packages, and unused
file compression utilities.}

\subsection{Combined effect of image update and  minification}

Package update and image minification remove different types of
vulnerabilities. The former is efficient against vulnerabilities that have
been fixed by package maintainers, while the latter targets unused
software. In two of the five tested images (images \textbf{S} and \textbf{U}), we find that combining update
and minification further reduces the number of vulnerabilities compared to
using only one of these processes
(Fig~\ref{fig:vulnerabilities}-\textbf{B}). \

\subsection{Differences between scanners}

\begin{figure}
  \centering

  \includegraphics[width=0.5\columnwidth]{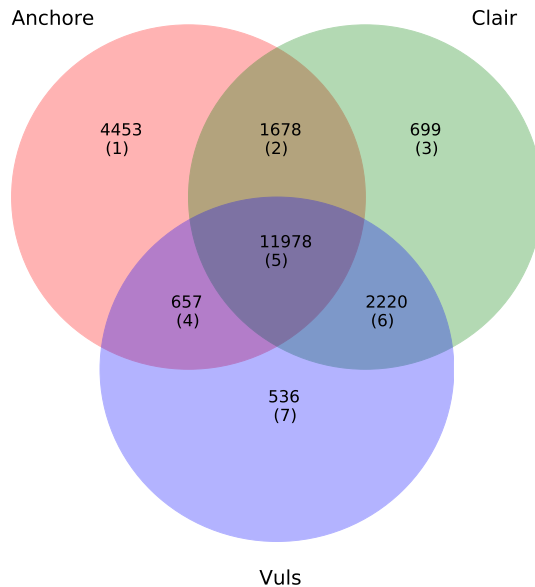}
  \caption{\label{fig:venn} Differences between vulnerabilities detected
by the different scanners. The Jaccard coefficients between the sets of
detected vulnerabilities are quite low, showing important discrepancies
between the scanners: Jaccard(Anchore, Clair) = 0.63, Jaccard(Anchore, Vuls) =
0.59, Jaccard(Vuls, Clair) = 0.80. Two Ubuntu 17.04 images weren't
included in this comparison as they cannot be scanned by Vuls.}
\end{figure}

The results presented so far were obtained with Anchore (Docker images) and
Stools (Singularity images). We scanned the Docker images with two other tools,
Clair and Vuls, to evaluate the stability of our results. Important
discrepancies were found between scanners (Fig~\ref{fig:venn}), in
particular between Anchore and the other two scanners, for which Jaccard
coefficients as low as 0.6 were found, meaning that scanning results only
overlapped by 60\%. Vuls and Clair appear to be in better agreement, with a
Jaccard coefficient of 0.8.

We analyzed these results and explained some reasons behind the observed
discrepancies. Out of 4453 vulnerabilities detected by Anchore only (region
\textbf{1} in Fig~\ref{fig:venn}), 4443 are found in the development
package of the C library (\texttt{linux-libc-dev} in Ubuntu and Debian).
Clair detects only Debian vulnerabilities in \texttt{linux-libc-dev},
whereas Vuls do not detect vulnerabilities in this package at all. Since Anchore
ignores Debian
vulnerabilities flagged as \texttt{minor}, it
might either detect (region \textbf{2}) or ignore (region \textbf{3})
the
Debian vulnerabilities detected by Clair in \texttt{linux-libc-dev}. The
remaining 10 vulnerabilities in region \textbf{1} are found in sub-packages
of vulnerable packages: they are correctly reported by Anchore and missed
by Vuls and Clair. 

Many vulnerabilities in region \textbf{3} and \textbf{4} are from images
based on Ubuntu 14.04. In the Ubuntu CVE tracker database used by Clair and
Anchore, there are two entries for Ubuntu 14.04: one for LTS (Long-Term
Support), a Ubuntu release with 5 years of technical support, and another one
for ESM (Extended Security Maintenance), a release that provides security
patches beyond the 5 years covered by LTS. Although all the scanned images
are LTS, Clair refers to the ESM database entry while Anchore and Vuls refer to the
LTS database entry. The vulnerabilities present in region \textbf{3} due to
this discrepancy are incorrectly missed by Anchore and Vuls: they have been
detected in ESM but were already present in LTS. The vulnerabilities in
region \textbf{4} are incorrectly missed by Clair: they have been fixed in
ESM but are still present in LTS. 

Some vulnerabilities in region \textbf{6} are due to bugs in Anchore: the
\textit{epoch bug} ignores vulnerabilities related to package versions that
contain an epoch (\texttt{:}); the \textit{out of standard bug} ignores
vulnerabilities that are ignored by the Ubuntu distribution. We reported
these bugs to the Anchore developers through their Slack channel. Some
vulnerabilities in region \textbf{6} are also due to the fact that Anchore
intentionally ignores Debian vulnerabilities flagged as \texttt{minor}.

Finally, 32 vulnerabilities that are flagged temporary by the Debian
distribution are reported by Vuls but not by Anchore or Clair (region
\textbf{7}). The remaining 504 vulnerabilities in this region are all found
in CentOS images. We weren't able to explain why they were detected by Vuls
only.

\section{Discussion}

There is a widespread issue with security vulnerabilities in container
images used for neuroimaging analyses, and it is likely to impact other
scientific disciplines as well. As shown in our results, it is common for container images
to hold hundreds of vulnerabilities, including several of critical
severity. Container images are impacted regardless of the type of analyses
that they support, and the main OS distributions Ubuntu, Debian and CentOS
are all affected.

Software updates remove about two-thirds of the vulnerabilities found and
should certainly be considered the primary solution to this problem.
However, in neuroimaging as in other disciplines, software updates are
generally discouraged because they can affect analysis results by
introducing numerical perturbations in the
computations~\cite{gronenschild2012effects,glatard2015reproducibility}. We
believe that this position is not viable from an IT security perspective,
and that it could endanger the entire Big Data processing infrastructure.
 Instead, we advocate a more systematic
analysis of the numerical schemes involved in data analyses, which, coupled
with software testing, would make the analyses robust to software updates.
As a first step, the packages impacting the analyses could be specifically
identified and the others updated, which would largely remove
vulnerabilities.

Ultimately, software updates should even occur at runtime rather than when
the container image is built. Indeed, it is likely that container images
used for scientific data analyses be built only occasionally, perhaps every
few weeks when a release becomes available, which may not be compatible
with the frequency of required security updates. In fact, there is no
definite reason for the application software release cycle to be
synchronized with security updates, and security updates shouldn't be
dependent on application software developers. Instead, we think it would be
relevant for analytics engines to (1) systematically apply security updates
when containers start, and (2) run software tests provided by application
developers, including numerical tests, before running analyses. 

Implementing such a workflow, however, requires a long-term endeavour to
evaluate broadly the stability of data analysis pipelines, and to develop
the associated software tests. For the shorter term, we identified the
following recommendations for application developers to reduce the number
of security vulnerabilities in container images:

\begin{enumerate}
\item \emph{Introduce software dependencies cautiously}. Software
dependencies come with a potential security toll that is often neglected.
For instance, it can be tempting to add a complete toolbox to implement a
relatively minor operation in a data analysis pipeline, such as a data
format conversion, while the same functionality might be available in the
existing dependencies of the pipeline, albeit in a less convenient way. \revision{Some package managers such as \texttt{apt} also 
support ``weak'' package dependencies that are recommended but not necessarily required for installation. Installation of such suggested packages 
can be avoided using specific options of the package manager (\texttt{--no-install-recommends} in the case of \texttt{apt}).}
\revision{Using lightweight base OS images such as Alpine Linux can also reduce the number of unnecessary dependencies. 
However, developers should ensure that this doesn't lead to installing 
extra dependencies without using the package manager, as mentioned in point (iv) below.}

\revision{\item \emph{Minify container images.} \revision{Minifying container images 
is another way to reduce software dependencies. However,} the automated minification
process that we used in our study is unwieldy for a routine use, as it
requires capturing execution traces with ReproZip to reconstruct the graph
of package dependencies required for the analysis. In practice, it would be
more practical for software developers to identify and remove unnecessary
dependencies when they build containers, based on their knowledge of the
application.}

\item \emph{Use OS releases with long-term support.} Security updates are
not provided for OS distributions that reached end of life. When a given
release of a data analysis pipeline is expected to be used over a long
period of time, typically several years as it is common in neurosciences,
the life cycle of the distribution release should be considered when
choosing a base container image. OS distributions have very different life
cycle durations, as long and short life cycles serve different purposes.
For instance, among RedHat-based distributions, Fedora release a new version
every 6 months and provide maintenance for about a year, while CentOS
release every 3-5 years and provide maintenance for 10 years. Similarly,
Ubuntu \revision{and Debian} LTS (long-term support) distributions provide security updates
for \revision{at least} 5 years. 

\item \emph{Install packages, not files}. \revision{Regardless of their support status, base OS distributions 
covered by the previous recommendation rarely include 
scientific software. Since 
 vulnerability scanners detect vulnerabilities from the list of installed packages}, vulnerabilities contained in
\revision{files installed without the package manger remain undetected.} \revision{To reduce such occurrences, scientific software 
should as much as possible be linked dynamically against libraries provided by the base OS distribution. The use of} domain-specific
\revision{repositories} such as \href{http://neuro.debian.net}{NeuroDebian} or
\href{https://docs.fedoraproject.org/en-US/neurofedora/overview/}{NeuroFedora}
in neuroimaging \revision{is also} useful in this respect\revision{, as
these distributions facilitate transparent software updates and might be covered by image scanners in the future}.

\item \emph{Run image scanners during continuous integration.} Scanning
container images can be a cumbersome process that
could be asynchronously executed during continuous integration (CI),
through tools such as Travis CI or Circle CI. Including security scans in
CI also allows developers to identify vulnerabilities quickly,
before new software versions are released. \revision{The Anchore documentation includes
specific instructions on how to do so at \url{https://anchore.com/cicd}. The Stools repository also includes 
a \href{https://github.com/singularityhub/stools-clair/blob/master/.travis.yml}{Travis CI file}
that can be reused for this purpose.}
\end{enumerate}

Describing specific attacks that would exploit
vulnerabilities in container images is out of the scope of our study. We
believe that such attacks are likely to exist \revision{given that critical vulnerabilities 
allowing for arbitrary code execution or file injection were found.}
Attacking
systems through containers remains challenging due to their relative isolation
from the host system. Under the assumption that \revision{container host}
users can be trusted, attackers would have to be remote,
either in the same network or on a remote network. Two main types of
attacks can be envisaged in these conditions: network-based attacks,
exploiting vulnerabilities in network clients installed in the container,
and data-based attacks, exploiting vulnerabilities through the processing
of malicious data injected through third-party systems.

Several types of escalation attacks could be envisaged once remote
attackers gain access to the container, in particular related to \revision{(1) stealing user credentials}, (2) using
the resources allocated to the container for malicious use, resulting in denial of service, and
(3) attacking a host network service, for instance \revision{a locally-accessible server} or a file
system daemon. Exploits in the host kernel to break out of the container
are always possible but unlikely assuming that the host system is
maintained by professional system administrators.

\section{Conclusion}

Most container images \revision{used in} scientific data analyses
contain hundreds of security vulnerabilities, many of which are critical.
In the short term, application software developers can address this issue
by: (1) reducing software dependencies, and (2) applying regular security
updates, which requires using OS distributions with long-term support.
Longer term, data analysis pipelines would benefit from in-depth stability
analysis, to ensure that analytical results are not affected by security
updates.

This conclusion is not an alarming message urging \revision{system} administrators  
to ban containers from their systems. User-controlled container images are
just one of many end-user artifacts that could serve as attack vectors, and
to our knowledge no attack has been described to exploit them. More
traditional types of attacks targeting user credentials or network
connections are likely to remain more common.

\section{Availability of Data and Materials}

The data and scripts used in this study are available on GitHub
with a Jupyter notebook to regenerate the figures:
\begin{enumerate}
\item Project name: container-vulnerabilities-paper
\item Project home page:\\ \url{https://github.com/big-data-lab-team/container-vulnerabilities-paper}
\item Operating system: Platform independent
\item Programming language: Python
\item Other requirements: Jupyter >= 1.0.0, Matplotlib >= 3.3.0, NumPy >= 1.19.1, Pandas >=1.0.5, SciPy >= 1.6.0
\item License: GNU General Public License v3.0
\end{enumerate}

\section{Availability of Supporting Data}

Kaur B; Dugré M; Hanna A; Glatard T: Supporting data for ``An Analysis of
Security Vulnerabilities in Container Images for Scientific Data Analysis"
GigaScience Database. 2021. \url{http://dx.doi.org/10.5524/100880}

\section{Competing interests}

The authors declare that they have no competing interests.

\bibliographystyle{abbrv}
\bibliography{bibliography}
\end{document}